\documentclass[letterpaper, 10 pt, conference]{ieeeconf} 
\IEEEoverridecommandlockouts 
\usepackage{graphics} 
\usepackage{epsfig} 
\usepackage{times, bm} 
\usepackage{amsmath} 
\usepackage{amssymb}
\usepackage{hyperref}
\usepackage[font=small,labelfont=bf]{caption}
\usepackage{cite, float}
\usepackage{dblfloatfix}

\title{\LARGE \bf
20-fold Accelerated 7T fMRI Using Referenceless Self-Supervised Deep Learning Reconstruction
}

\author{Omer Burak Demirel$^{1,2}$,
        Burhaneddin Yaman$^{1,2}$,
        Logan Dowdle$^{2,3}$,
        Steen Moeller$^{2}$,
        Luca Vizioli$^{2,3}$,\\
        Essa Yacoub$^{2}$,
        John Strupp$^{2}$,
        Cheryl A. Olman$^{2}$,
        K\^{a}mil U\u{g}urbil$^{2}$
        and Mehmet Ak\c{c}akaya$^{1,2}$
        \vspace{-0.4cm}
\thanks{$^{1}$Department of Electrical and Computer Engineering and, $^{2}$Center for Magnetic Resonance Research and, $^{3}$Department of Neurosurgery, University of Minnesota, Minneapolis, MN, USA.  {\tt\small e-mails: \{demir035, yaman013, dowdl016, moell018, lvizioli, yaco0006, strupp, caolman, ugurb001, akcakaya\}@umn.edu}}}

\begin{document}
\maketitle
\thispagestyle{empty}
\pagestyle{empty}

\begin{abstract}
High spatial and temporal resolution across the whole brain is essential to accurately resolve neural activities in fMRI. Therefore, accelerated imaging techniques target improved coverage with high spatio-temporal resolution. Simultaneous multi-slice (SMS) imaging combined with in-plane acceleration are used in large studies that involve ultrahigh field fMRI, such as the Human Connectome Project. However, for even higher acceleration rates, these methods cannot be reliably utilized due to aliasing and noise artifacts. Deep learning (DL) reconstruction techniques have recently gained substantial interest for improving highly-accelerated MRI. Supervised learning of DL reconstructions generally requires fully-sampled training datasets, which is not available for high-resolution fMRI studies. To tackle this challenge, self-supervised learning has been proposed for training of DL reconstruction with only undersampled datasets, showing similar performance to supervised learning. In this study, we utilize a self-supervised physics-guided DL reconstruction on a 5-fold SMS and 4-fold in-plane accelerated 7T fMRI data. Our results show that our self-supervised DL reconstruction produce high-quality images at this 20-fold acceleration, substantially improving on existing methods, while showing similar functional precision and temporal effects in the subsequent analysis compared to a standard 10-fold accelerated acquisition.
\newline

\end{abstract}

\section{INTRODUCTION}
Functional MRI (fMRI) has been crucial in expanding our understanding of human perception and cognition \cite{logothetis2008we}. Even though significant progress has been made in fMRI acquisition and reconstruction \cite{uugurbil2013pushing}, additional improvements for coverage and resolution are imperative to contribute to our broader understanding of the brain function. 

As fMRI acquisitions have become a standard part of large-scale studies, such as the Human Connectome Project (HCP) \cite{van2013wu}, several accelerated MRI methods have been utilized for improved resolution and coverage. Simultaneous multi-slice (SMS) imaging is commonly used in functional neuroimaging with its ability for rapid high-resolution with whole brain coverage \cite{moeller2010multiband,setsompop2012blipped}. SMS imaging is typically further combined with in-plane acceleration for improving the resolution of fMRI at ultrahigh field strengths, while maintaining a reasonable echo time \cite{uugurbil2013pushing}. These accelerated acquisitions are then reconstructed using linear SMS and parallel imaging reconstruction \cite{setsompop2012blipped,cauley2014interslice,moeller2010multiband}.
However such reconstructions suffer from spatially varying noise amplification \cite{hamilton2017recent,barth2016simultaneous}, and may suffer from inter and intra-slice residual aliasing artifacts \cite{todd2016evaluation} which will further be exacerbated when operating at higher acceleration rates.

Recently, deep learning (DL) methods have gained immense interest as an alternative to improve accelerated MRI \cite{knoll2020deep,hammernik2018learning,schlemper2017deep,aggarwal2018modl}. In particular, physics-guided DL methods have emerged as a powerful strategy due to their robustness and improved reconstruction quality \cite{hammernik2018learning,schlemper2017deep,aggarwal2018modl,hosseini2020dense}, showing promising results at higher acceleration rates where conventional methods struggle to maintain high quality reconstruction. In general, DL reconstruction methods are trained in a supervised manner, using fully-sampled data as reference \cite{knoll2020deep,hammernik2018learning,schlemper2017deep,aggarwal2018modl}. However, such reference data is not available in highly-accelerated fMRI acquisitions. Recently, self-supervised learning has been proposed for training physics-guided DL reconstruction using only undersampled data, showing similar performance to supervised learning \cite{yaman2020self}.

In this work, we adapt a recent self-supervised DL \cite{yaman2020self} strategy to reconstruct an SMS-accelerated HCP-style retinotopic mapping acquisition and show that the results of subsequent fMRI analysis are unaltered by the use of such a regularized non-linear DL reconstruction. In doing so, we establish the feasibility of reconstructing 20-fold accelerated 7T fMRI data using self-supervised DL reconstruction. Results show that the proposed DL reconstruction substantially improves upon conventional methods at the 20-fold acceleration rate, while maintaining comparable signal fidelity to conventional methods at standard 10-fold acceleration for fMRI analysis.

\section{METHODS}

\begin{figure*}[t]
 \begin{center}
          \includegraphics[trim={0 0 0 0},clip, width=6.2 in]{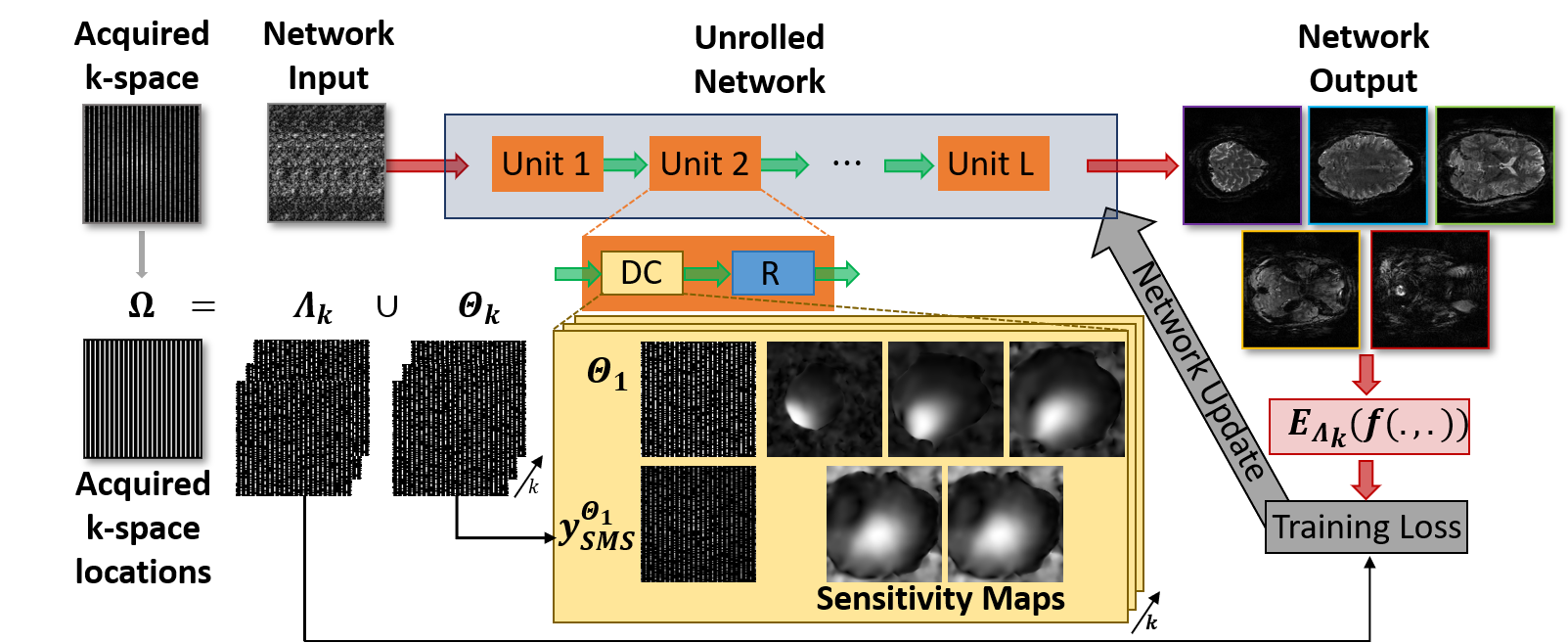}
     \end{center}
      \vspace{-.25cm}
  	\caption{A schematic of the self-supervised approach used in this study that does not require fully-sampled data. The iterative algorithm in Equations (\ref{Req})-(\ref{DCeq}) is unrolled for a fixed number of iterations. Each iteration consists of a regularizer unit which is implemented by a ResNet, and a data consistency unit that enforces fidelity with the acquired k-space. Self-supervised training splits the acquired samples into $K$ pairs of disjoint sets ($\Omega = \Theta_k \cup \Lambda_k$), where $\Theta_k$ is used for data-consistency and $\Lambda_k$ is used for training loss. This strategy allows a training without fully-sampled data, which has been a main hindrance for the adaptation of DL reconstruction for highly-accelerated fMRI. Multi-mask SSDU is utilized with $K$= 6.}
  	\label{fig:1}
  	\vspace{-.35cm}
\end{figure*}

\subsection{Self-supervised DL Reconstruction for SMS fMRI} \label{ssdu}

The forward model for SMS acquisition is given as: 

\begin{equation}
    \mathbf{y}_{SMS}^{\Omega} =  \sum _{i=1}^{S}\mathbf{E}_{i}^{\Omega}\mathbf{x}_{i} + {\bf n},
\end{equation}
where $\bf{y}_{SMS}^\Omega \in {\mathbb C}^P$ is the acquired SMS k-space, $\Omega$ is the in-plane undersampling pattern, $S$ is number of simultaneously excited slices, ${\bf x}^{i} \in {\mathbb C}^{M \times N}$ is the $i^{th}$ slice image, $\bf{E}_i^{\Omega}: {\mathbb C}^{M \times N} \to  {\mathbb C}^P$ is the multi-coil encoding operator of the $i^{th}$ slice, and ${\bf n}$ is measurement noise. The notation can be condensed by letting ${\bf x}_{SMS}$ be the concatenation of the individual slices $\{{\bf x}_i\}_{i=1}^S$ along the readout direction \cite{moeller2010multiband,demirel2021improved}, and $\mathbf{E}_{SMS}^{\Omega} = [\mathbf{E}_{1}^{\Omega} \dots \mathbf{E}_{S}^{\Omega}]$, which yields
\begin{equation}
    \mathbf{y}_{SMS}^{\Omega} =  \mathbf{E}_{SMS}^{\Omega}\mathbf{x}_{SMS} + {\bf n}.
\end{equation}
For high acceleration rates, this system is generally ill-conditioned. Thus, a regularized least squares formulation may be used for the inverse problem: 
\begin{equation} \label{Eq:recons1}
    \arg \min_{{\bf x}_{SMS}} ||\mathbf{y}_{SMS}^{\Omega} -  \mathbf{E}_{SMS}^{\Omega}\mathbf{x}_{SMS}||_2^2 + {\cal R}(\mathbf{x}_{SMS}),
\end{equation}
where the first term enforces data consistency (DC) with the acquired measurements and $\cal{R}(\cdot)$ is a regularizer. This least squares problem is typically solved by using iterative techniques that decouples the DC term and the regularizer into a series of sub-problems \cite{fessler2020optimization}. One such solution is variable splitting with quadratic penalty \cite{aggarwal2018modl,yaman2020self}: 
\begin{align}
\mathbf{z}_{SMS}^{(l-1)} &= \arg \min_{\mathbf{z}_{SMS}} \mu\|  \mathbf{x}_{SMS}^{(l-1)} - \mathbf{z}_{SMS} \|_2^2 + {\cal{R}}(\mathbf{z}_{SMS}), \label{Req} \\
    \mathbf{x}_{SMS}^{(l)} &= \arg \min_{\mathbf{x}_{SMS}} \|\mathbf{y}^{\Omega}_{SMS}-\mathbf{E}_{SMS}^{\Omega}\mathbf{x}_{SMS}\|^2_2 \nonumber \\ &\quad\quad\quad\quad\quad+ \mu \| \mathbf{x}_{SMS} - \mathbf{z}_{SMS}^{(l-1)} \|_2^2, \label{DCeq}
\end{align}
where $\mathbf{x}_{SMS}^{(l)}$ is the reconstructed slices at iteration $l$, while $\mathbf{z}_{SMS}^{(l)}$ is an auxiliary variable and $\mu$ is the penalty parameter.  This iterative algorithm is unrolled for a fixed number of iterations in physics-guided DL reconstruction \cite{knoll2020deep}, and sub-problem (\ref{Req}) is solved implicitly using neural networks, while (\ref{DCeq}) is solved by conjugate gradient \cite{aggarwal2018modl}. The network output for a given test dataset can be written as $f(\mathbf{y}_{SMS}^{\Omega},\mathbf{E}_{SMS}^{\Omega};\bm{\theta})$, where ${\bm \theta}$ are the learnable parameters of the unrolled network.

Although physics-guided neural networks are conventionally trained in a supervised manner, the lack of reference training data for high-resolution fMRI studies prevents this strategy. However, a recent work entitled Self-Supervised learning via Data Undersampling (SSDU) has shown similar performance to supervised learning by splitting the acquired locations, $\Omega$, into two disjoint sets, $\Theta$ and $\Omega$, where $\Theta$ is used in the DC units of the unrolled network and $\Lambda$ is used to define the k-space loss \cite{yaman2020self}. A multi-mask version of SSDU has also been proposed to improve reconstruction quality at very high acceleration rates by utilizing these disjoints sets multiple times \cite{yaman2020multi,yaman2020multiisbi}. In this study,  multi-mask SSDU is used with the following loss function:  

\begin{equation}
    \min_{\bm{\theta}} \frac{1}{N \cdot K}\sum_{n=1}^{N}\sum_{k=1}^{K}\mathcal{L}(\mathbf{y}_{SMS}^{\Lambda_k,n},\mathbf{E}_{SMS}^{\Lambda_k,n}(f(\mathbf{y}_{SMS}^{\Theta_k,n},\mathbf{E}_{SMS}^{\Theta_k,n};\bm{\theta}))),
\end{equation}
where $N$ is the number of training data in the database, $K$ is the number of multi-maks ($\Omega = \Theta_k \cup \Lambda_k$, $k \in {1,\cdots,K}$), $\mathcal{L}(\cdot,\cdot\cdot)$ is the k-space loss between the unseen acquired points and the network output and the network is parametrized by ${\bm \theta}$. The $k^{th}$ training and loss masks used on the $n^{th}$ training data sample are represented as $\Lambda_k,n$ and $\Theta_k,n$, respectively. A schematic of the implementation is shown in Figure \ref{fig:1}. 

\begin{figure}[t]
 \begin{center}
          \includegraphics[trim={0 0 0 0},clip, width=0.98\columnwidth]{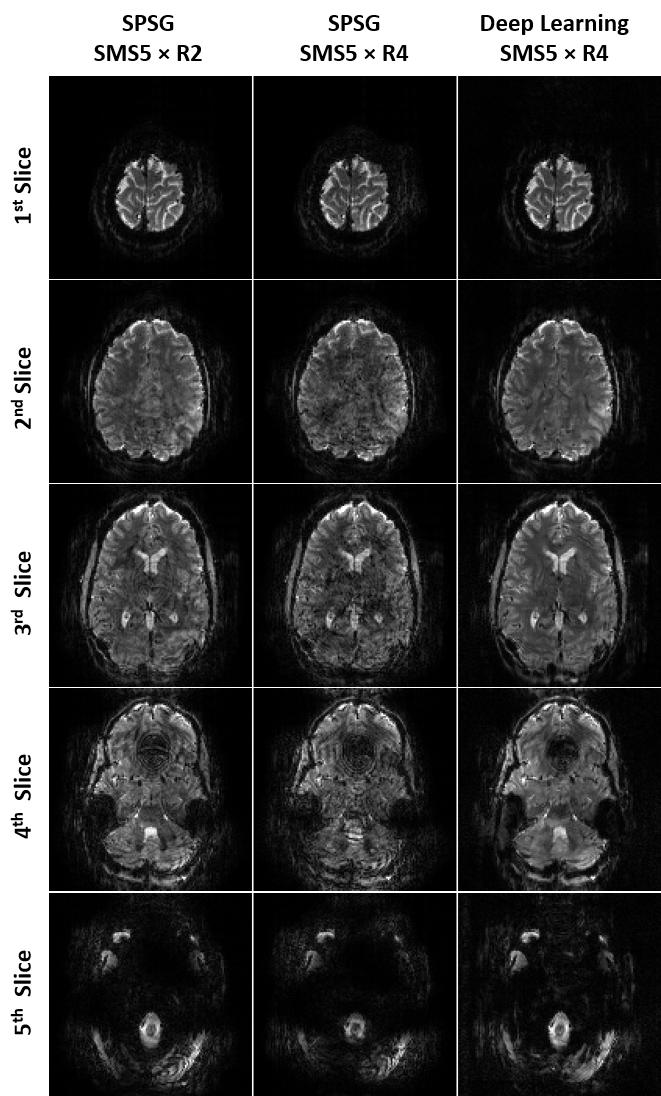}
     \end{center}
      \vspace{-.25cm}
  	\caption{Representative slices from the HCP-style retinotopy acquisition, reconstructed with SPSG at the standard 10-fold acceleration, and with SPSG and proposed self-supervised DL at a retrospective 20-fold acceleration. Baseline image quality is based on the 10-fold SPSG reconstruction. At 20-fold acceleration, SPSG suffers from visible noise amplification and residual artifacts, while the self-supervised DL reconstruction substantially suppresses these across all the slices shown.
  	}
  	\label{fig:2}
  	\vspace{-.35cm}
\end{figure}

\subsection{Imaging Experiments} 
Imaging was performed on a 7T Siemens Magnex Scientific (Siemens Healthineers, Erlangen, Germany) system with a 32-channel receiver head coil-array. 8 subjects were imaged using the HCP-retinotopy protocol detailed in \cite{benson2018human}. The study was approved by our institutional review board, and written informed consent was obtained before each examination.

For each subject, six 5-minute acquisitions were obtained with the protocol in \cite{benson2018human}. In this study, the first 2 experimental runs, which are the 
rotating wedge (RETCCW and RETCW) retinotopic mapping paradigms, were utilized. RETCCW was collected with an anterior to posterior phase encoding direction, while RETCW was collected with a posterior to anterior phase encoding direction. The following imaging parameters were used similar to the 7T HCP fMRI protocol \cite{uugurbil2013pushing}: SMS factor = 5, in-plane acceleration = 2, resolution: 1.6mm isotropic, TR = 1s and whole-brain coverage. 
In both runs, the subjects viewed wedges which completed a rotation every 32 seconds (0.3125 Hz). These data were used as the retinotopic representation within the visual cortex.

The standard acquisition at 5-fold SMS and 2-fold in-plane acceleration (overall 10-fold acceleration), reconstructed with conventional split-slice GRAPPA (SPSG)  \cite{cauley2014interslice} was used as the baseline for image quality and fMRI analyses. These acquired k-space data were further retrospectively undersampled to 5-fold SMS and 4-fold in-plane acceleration (overall 20-fold acceleration), which were then reconstructed using the proposed self-supervised DL reconstruction, as well as conventional SPSG. In the following, SMS5$\times$R2 (as 10-fold acceleration) notation is used for the standard acquisition, while SMS5$\times$R4 (as 20-fold acceleration) notation is used for the retrospectively sub-sampled datasets.

\subsection{Functional Processing and Retinotopic Analyses}
The reconstructions were each minimally processed identically using \texttt{afni\_proc.py}, a tool available with AFNI \cite{cox1996afni}. To account for differences from the different phase encoding directions, distortion correction was performed. Subsequently, motion correction was performed using the distortion-corrected median image as a registration target. These distortion- and motion-corrected data were then scaled such that each voxel had a mean of 100. The processed data were then imported into MATLAB. The clockwise wedge data were time-reversed and both datasets were shifted to align the hemodynamic responses for identical wedge positions. Next, 9 nuisance regressors were projected out from each dataset. These were polynomials up to order 3 and the 6 (rigid-body) motion estimates from the motion-correction step. The mean of the two datasets were calculated and an FFT was performed to determine the amplitude and phase at 0.3125Hz. In order to threshold the data, the coherence between the two runs was also calculated using the MATLAB tool \texttt{mscohere}. The phase maps were compared for each reconstruction using a coherence threshold of 0.4.

A region of interest was then created from the baseline SMS5$\times$R2 data reconstructed with SPSG using this threshold. A number of metrics were calculated for this region. This included the average absolute phase error, relative to the original SMS5$\times$R2 SPSG reconstruction, of the deep learning and SPSG reconstructions of the SMS5$\times$R4 accelerated data. Additionally, temporal signal to noise ratio (tSNR) was calculated for each reconstruction, estimated as the mean of the signal after processing divided by the standard deviation of the residual following the removal of nuisance regressors. Circular statistics were handled using CircStat \cite{berens2009circstat}.

\subsection{Implementation Details}

Multi-mask SSDU \cite{yaman2020self,yaman2020multi,yaman2020multiisbi} was used with $K$ = 6 masks to train a physics-guided DL reconstruction network. The sub-problems in (\ref{Req}) and (\ref{DCeq}) were unrolled for 10 iterations. Each iteration contains a DC unit and a regularizer. The former was solved by 10 conjugate gradient steps \cite{aggarwal2018modl}, while the latter was solved by a convolutional neural network that utilizes a ResNet structure \cite{timofte2017ntire}. ESPIRiT was used to generate sensitivity maps from low-resolution calibration scans \cite{uecker2014espirit}. The network was trained using Adam optimizer with an $\ell_1$-$\ell_2$ loss in k-space \cite{yaman2020self} with a learning rate of $3\cdot10^{-4}$ over 80 epochs. The network had a total of 592,129 trainable parameters.

For SMS imaging, to avoid boundary artifacts during regularization, the CAIPI \cite{setsompop2012blipped} shifts were removed to position each slice to the middle of the FOV before the regularizer ResNet unit, and re-arranged back to the acquired CAIPI shifts before the DC units. The ResNet unit operated on a 2-channel input (real and imaginary), and the SMS = 5 slices were concatenated along the readout direction, as detailed in Section \ref{ssdu}.

Training was performed on 17 slab groups from 4 subjects. Only one time-frame per subject was used. Therefore no temporal correlations/redundancies were exploited by the DL reconstruction. Testing was performed on the whole volume and whole time series of 4 subjects unseen by the network. All training was performed using TensorFlow in Python.

\section{RESULTS}

SPSG reconstruction at SMS5$\times$R2 is shown in Figure \ref{fig:2} (first column) as the baseline image quality. For the SMS5$\times$R4 accelerated data reconstructed with both SPSG and self-supervised DL, visible noise reduction is observed with the proposed self-supervised DL reconstruction. Compared to SPSG reconstruction at 20-fold acceleration, self-supervised DL removes residual aliasing artifacts leading to a closer match to the reference image quality. Substantial improvement is observed in thalamus region (third slice) with the proposed DL method.

\begin{figure}[t]
 \begin{center}
          \includegraphics[trim={0 0.3cm 0 0},clip, width=\columnwidth]{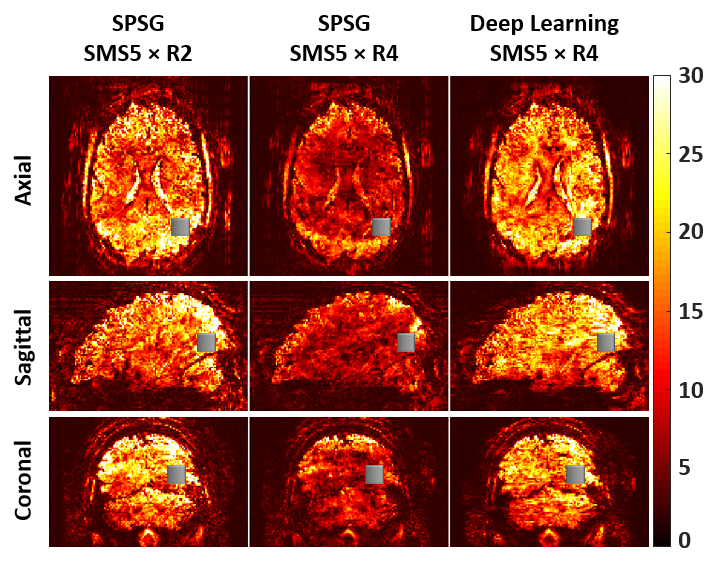}
     \end{center}
      \vspace{-.25cm}
  	\caption{tSNR maps of axial, sagittal and coronal slices using SPSG at standard 10-fold acceleration, and using SPSG and proposed self-supervied DL at retrospective 20-fold acceleration. Substantial tSNR gains are observed in the self-supervised DL reconstruction compared to SPSG. The mean and standard deviation of tSNR values in the grey boxes are 18.99$\pm$0.32, 8.94$\pm$0.25 and 17.38$\pm$0.33 for SMS5$\times$R2 SPSG, SMS5$\times$R4 SPSG and SMS5$\times$R4 self-supervised DL, respectively, confirming these observations. 
  }
  	\label{fig:3}
  	\vspace{-.35cm}
\end{figure}

\begin{figure}[t]
 \begin{center}
          \includegraphics[trim={0 0 0 0},clip, width=0.96\columnwidth]{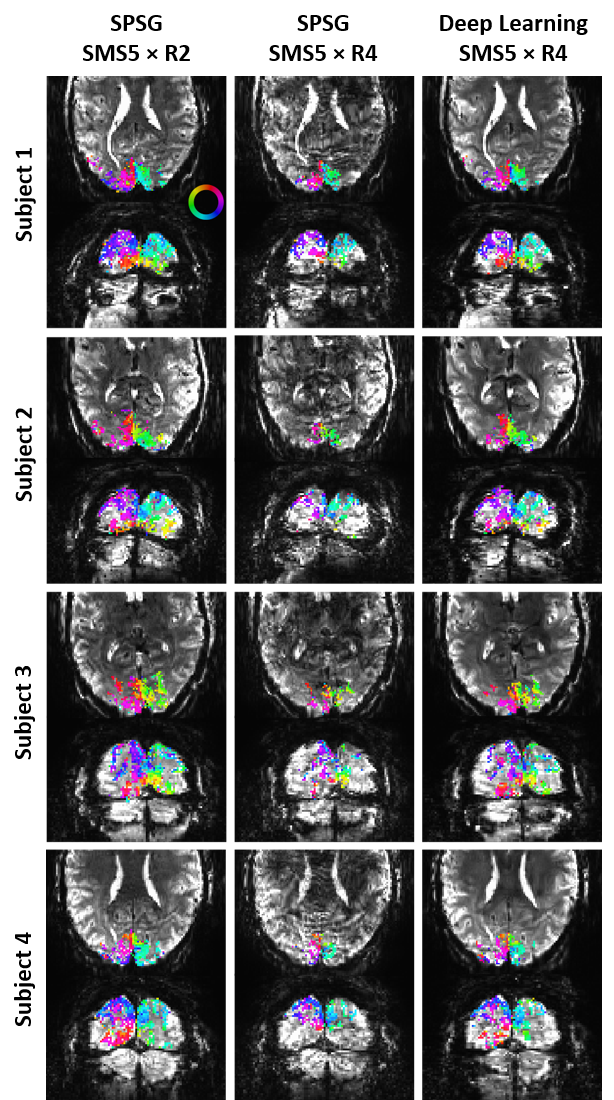}
     \end{center}
      \vspace{-.25cm}
  	\caption{Phase maps for each reconstruction type with a coherence threshold of 0.4, limited to clusters greater than 40 voxels. Color indicates polar angle, in radians (please see the legend in top left figure). Right side of the brain is on right.}
  	\label{fig:4}
  	\vspace{-.35cm}
\end{figure}

Fig. 3 depicts tSNR maps of axial, sagittal and coronal slices of a subject reconstructed with SPSG and self-supervised DL reconstructions prior to any processing for distortion or motion. The first column depicts the baseline 10-fold accelerated acquisition using SPSG reconstruction. Self-supervised DL reconstruction substantially improves upon SPSG at SMS5$\times$R4 acceleration showing good fit to reference tSNR maps. Mean and standard deviation of the tSNR values in a $10\times10\times10$ voxel (gray boxes) agree with these visual observations. 10-fold acceleration has the highest tSNR value (18.99$\pm$0.32) while SPSG at SMS5$\times$R4 acceleration has the lowest tSNR value (8.94$\pm$0.25). Self-supervised DL reconstruction at SMS5$\times$R4 acceleration leads to a substantially improved tSNR value (17.38$\pm$0.33) close to baseline tSNR of the 10-fold accelerated acquisition.

Following distortion and motion correction, the average of the absolute difference in phase estimates across subjects was 0.36$\pm$0.02 radians for the SMS5$\times$R4 self-supervised DL and 0.46$\pm$0.40 radians for the SMS5$\times$R4 SPSG reconstructions, within the ROI derived from the original SMS5$\times$R2 SPSG reconstruction. In this setting, the average tSNR within the ROI was 32.5$\pm$3.2 for SMS5$\times$R2 SPSG, 22.5$\pm$2.0 for SMS5$\times$R4 self-supervised DL and 15.1$\pm$1.5 for the SMS5$\times$R4 SPSG reconstructions.

\section{Discussion}

We proposed a self-supervised physics-guided deep learning reconstruction for highly-accelerated HCP-style fMRI. A retrospective study with SMS5$\times$R4 acceleration was evaluated both for image reconstruction and subsequent fMRI analysis. Proposed self-supervised DL reconstruction showed improved image quality compared to existing SPSG reconstruction. tSNR improvement is observed with self-supervised DL reconstruction compared to SPSG. Furthermore, fMRI analysis shows that self-supervised DL reconstruction has higher coherence than the SPSG reconstruction in the visual cortex similar to standard acquisition.

For the 20-fold accelerated data, the self-supervised DL reconstruction strongly outperforms the SPSG reconstruction across multiple metrics, including tSNR and absolute difference in phase estimates. For the latter, substantial improvement in standard deviation is observed in addition to a 20.40$\pm$0.96\% improvement in the mean, highlighting less extreme errors in phase estimations. This is also apparent from the images themselves, which possess visible artifacts for SMS5$\times$R4 SPSG reconstruction. With the DL reconstruction, these artifacts are no longer visible, and the image structure resembles the original SMS5$\times$R2 data. The self-supervised DL functional data are also associated with higher tSNR compared to the SMS5$\times$R4 SPSG reconstruction.

While the SMS5$\times$R4 DL reconstruction do not fully reproduce the SMS5$\times$R2 case, the phase maps in Figure \ref{fig:4} suggest that much of the important temporal information is retained across the two runs. The activation patterns are robust, and retinotopic mapping is preserved, leading to highly correlated phase values and low error in phase estimation. 

\section{Conclusion}
The proposed self-supervised DL reconstruction reduced residual artifacts and improved the tSNR compared to existing reconstruction technique at 20-fold accelerated fMRI acquisition. The fMRI analysis led to similar results compared to standard acquisition, showing that self-supervised DL reconstruction did not alter the temporal effects or functional spatial precision.

\section*{ACKNOWLEDGMENTS}
First two authors contributed equally to this work. Grant support: NIH, Grant numbers: P30NS076408, R01HL153146, U01EB025144, P41EB027061; NSF, Grant number: CAREER CCF-1651825.

\bibliographystyle{IEEEbib}
\bibliography{reference}

\end{document}